\documentclass[alpha-refs]{wiley-article} % blind

\usepackage{color}
\usepackage{ulem}

\usepackage{siunitx}
\usepackage{algorithm}
\usepackage{algorithmicx}
\usepackage{algpseudocode}
\usepackage{amsmath,amssymb,amsfonts}
\usepackage[title]{appendix}
\usepackage{bm}
\usepackage{booktabs}
\usepackage{cancel}
\usepackage{graphicx}
\usepackage{listings}
\usepackage{manyfoot}
\usepackage{mathrsfs}
\usepackage{multirow}
\usepackage{natbib}
\usepackage{textcomp}
\usepackage{xcolor}

\title{Multivariate Shared Frailty Cure-Rate models: a focus on Breast Cancer family history}

\author[1\authfn{1}]{Maria Veronica Vinattieri}
\author[2]{Marco Bonetti}
\author[3]{Kamila Czene}

\affil[1]{Department of Decision Sciences, Bocconi University, Via Guglielmo Röntgen 1, Milan, Italy}
\affil[2]{Carlo F. Dondena Research Center, Bocconi Institute of Data Science and Analytics, Department of Social and Political Sciences, Bocconi University, Via Guglielmo Röntgen 1, Milan, Italy}
\affil[3]{Department of Medical Epidemiology and Biostatistics, Karolinska Institutet, Nobels Väg 12A, Stockholm, Sweden}

\corraddress{Marco Bonetti, Department of Social and Political Sciences, Bocconi University, Via Guglielmo Röntgen 1, Milan, Italy}
\corremail{marco.bonetti@unibocconi.it}

\presentadd[\authfn{1}]{Department of Medical Epidemiology and Biostatistics, Karolinska Institutet, Nobels Väg 12A, Stockholm, Sweden}

\fundinginfo{Kamila Czene, Department of Medical Epidemiology and biostatistics, Karolinska Institutet, Grant Number: 2022-00584; Marco Bonetti, Department of Social and Political Sciences, Bocconi, Grant: European Union – Next Generation EU, under NRRP Mission 4 Component 2 Investment Line 1.5.}

\runningauthor{Vinattieri et al.}
\begin{document}
\maketitle
\begin{abstract}
We discuss a shift in perspective from traditional approaches to breast cancer risk prediction: modelling families rather than individuals as unit of analysis. By investigating the latent familial risk underlying breast cancer diagnoses, we introduce a Multivariate Shared Frailty Cure-Rate model. This model captures the familial risk as a shared frailty among members and explicitly accounts for a fraction of women not susceptible to breast cancer. We aim at identifying the high-risk families to better target screening and prevention, ultimately improving early detection. A comparative analysis with Cox models and univariate models—where a binary risk indicator acts as ``best guess'' for the latent high-risk group—is conducted using simulation studies and data from the Swedish Multi-Generational Breast Cancer registry. 

\noindent
We demonstrate the critical importance of using complete family history of breast cancer to accurately identify high-risk families and show that the Multivariate Shared Frailty Cure- Rate model, capturing both the fraction of non-susceptible subjects and the survival distribution among susceptibles, enhances explanatory power, improves prediction accuracy, and offers a broader representation of the disease process.

\end{abstract}

\section{Introduction}
Individual breast cancer risk prediction enables the design of targeted prevention strategies, potentially improving opportunities for early detection and treatment success. Numerous breast cancer risk prediction models have been extensively developed, incorporating several subject-specific risk factors to enhance their predictive power. 

\cite{gail1989projecting} estimated the long-term risk of developing breast cancer using a set of demographic and clinical individual risk factors such as the number of first-degree relatives (mother, sisters, and daughters) who have been diagnosed with breast cancer, age, reproductive history, and previous biopsy results. This model provides a 
\\
comprehensive risk assessment for women, often used in clinical settings to guide decisions regarding screening and preventive measures.

\cite{rosner2008risk} expanded upon the Gail framework by incorporating a broader set of reproductive covariates. This includes factors such as age at menarche, age at first childbirth, and age at menopause, all of which play significant roles in breast cancer risk. This model is also adjusted for time-dependent exposures, recognizing that risk changes over a woman's lifetime as her reproductive history and lifestyle evolve.

\cite{tyrer2004breast} introduced an even more comprehensive approach, integrating not only reproductive and familial factors but also genetic predispositions, such as gene mutations BRCA1 and BRCA2. This model also accounts for mammographic breast density, a key risk factor whose high values are associated with increased breast cancer risk. Body mass index and lifestyle factors, including alcohol consumption and physical activity, are also included as predictors, acknowledging the role that both genetic and environmental factors play in breast cancer disease onset (or detection).

BOADICEA (see for instance, \cite{antoniou2002comprehensive, antoniou2004boadicea, antoniou2008boadicea, thomas2004statistical}) model is a highly effective tool for breast cancer risk prediction and can be considered the gold standard for breast cancer risk assessment. It incorporates detailed genetic information, including BRCA1, BRCA2, PALB2, CHEK2, and ATM mutations, alongside polygenic risk scores to capture a broader range of genetic predispositions. This model integrates both family history and genetic data, complementing traditional models. Although it represents a significant step forward in individual breast cancer risk prediction, relying on detailed genetic data may restrict its application in clinical settings where such information is not accessible.

A significant limitation of all of these models, which we will refer to as ``univariate models'', is their emphasis on individual subjects as unit of analysis. While they can incorporate various covariates, including family history, we believe that they do not adequately account for the underlying familial structure that may contain additional information on risk profiles and survival dynamics. In contrast, we define ``multivariate models'' those models resulting from shifting from subjects to families as unit of analysis, recognizing that family members often share genetic and environmental risk factors.

Shared frailty models are a natural modelling option for analysing familial risk with multivariate survival data. These models allow for the inclusion of unobserved heterogeneity (frailty) that varies across families as, e.g., shown by \cite{yashin1995genetic} in a survival study among Danish twins. Similarly, \cite{hsu2006multivariate} demonstrated that frailty models are appropriate tools for handling familial correlations among survival times, potentially improving risk assessment. Without explicitly modelling the correlation among survival times of related subjects, an alternative is given by \cite{chen2002bayesian} who estimated survival times allowing for the incorporation of prior knowledge on the frailty distribution, within a Bayesian framework. 

We aim to contribute to breast cancer risk prediction modelling by combining a multivariate shared frailty model, which effectively captures the familial risk of breast cancer detection through the frailty component, with a cure-rate structure on ``non-susceptible'' subjects to breast cancer into what we called the Multivariate Shared Frailty Cure-Rate survival model. The fraction of non-susceptibles is the proportion of women who will never experience breast cancer detection, no matter how long they will live, while the subjects at risk are called ``susceptibles''. Incorporating the fraction of non-susceptibles offers greater flexibility compared to conventional survival models which assume that all women will eventually experience the event. This aligns with \cite{gorfine2007prospective}, who demonstrated the benefits of incorporating a cured fraction in survival models for better risk prediction. We make the probability of being non-susceptible and the survival function of susceptibles depend on the familial breast cancer frailty risk, similarly to what \cite{yu2008mixture} did on subject-specific covariates. 

% The effectiveness of the Multivariate Shared Frailty Cure-Rate survival model is evaluated through a comparative analysis. Enhancing risk prediction accuracy, explaining the magnitude of breast cancer detection by directly estimating the fraction of non-susceptibles, and providing insights into the distribution of breast cancer detection times among susceptibles are all part of its benefits.

In Section \ref{section:2} we introduce the Multivariate Shared Frailty Cure-Rate model and its key components. Section \ref{section:3} describes the simulation study designed to assess the performance of the proposed model through a comparative analysis. We then apply the models to the Swedish Multi-Generational Breast Cancer registry in Section \ref{section:4}. Finally, Section \ref{section:5} provides a discussion of the findings, limitations, and potential future directions.

\section{Model Specification}\label{section:2}
In this section, we present a detailed overview of the development of the Multivariate Shared Frailty Cure-Rate model with Lehmann structure. We describe the key assumptions, modelling framework, and mathematical formulation at the basis of the approach. This provides the foundation for both the simulation study and the real data application presented in the subsequent Sections.

\subsection{Multivariate Shared Frailty Cure-Rate model with Lehmann structure }\label{section:2.1}
Consider the families identified with $i=1,\dots,n$.
The observed survival data is $\underline X=\left(\underline X_1,\dots,\underline X_n\right)^T$ where, for the $i$th family, $\underline X_i=(\underline x_{i1},\dots, \underline x_{in_i})^T$ with family size (of female members) equal to $n_i$. For each family, we randomly identify among sisters one subject to be what we call the ``main subject'', whose survival data is $\underline x_{i1}$. The remaining $(n_i-1)$ members are consequently identified (in random order) as mother and sisters of the main subject. The distinction between mother and sisters is not strictly needed here, as we assume that their survival distributions are equal. This can be easily extended to account for dependence on each member's birth cohort.

For a generic subject, the survival data is the pair $\underline x=(x=\text{min}(t, c),\ \delta=\mathbb{I}(t \le c))^T$ where $t$ indicates the survival time, and $c$ indicates the administrative (independent) censoring time, both measured from the same origin which, in our case, is birth.

We model the age (in years) $T$ at breast cancer detection, where the family-specific conditional hazard function is denoted by $\lambda(t\mid r)$, with $T = t$ and the familial risk random variable $R = r$. The univariate multiplicative frailty model allows the hazard function to depend on the frailty quantity $R$ to capture the unobserved heterogeneity among subjects 
 (see for instance, \cite{duchateau2008frailty, ripatti2000estimation}): \[ \lambda(t\mid r) = r \ \lambda_0(t),\]
following the Proportional Hazards (PH) assumption (see, \cite{cox1972regression}) or \cite{lehmann1953power} structure, with $\lambda_0(t)$ the baseline hazard function. These assumptions are identical for proper survival functions and correspond to defining the conditional survival function $S(t\mid r)$ as \[S(t\mid r) = S_r(t) = [S_0(t)]^r. \] 

We extend the Lehmann structure to the model defined as the more general Lehmann Cure-Rate (LCR) model by applying the power transformation to a Cure-Rate (CR) survival function: \begin{eqnarray}
\label{formula:1}
    &\text{CR}: \ S_0(t) = p + (1-p) \widetilde{S}(t),
    &\text{LCR}: \ S_r(t) = \left[ p + (1-p) \widetilde{S}(t) \right]^r, \ r >0,
\end{eqnarray} with $\widetilde{S}(t)$ a proper survival function (with corresponding density function $\widetilde{f}(t)$) which describes the time-to-event distribution of the $(1-p)$ fraction of susceptibles. The Lehmann structure allows the proportion of non-susceptibles and the survival function among susceptibles to vary with familial frailty risk, thus capturing family-specific heterogeneity in both disease susceptibility and progression. In (non-web) Appendix A we comment on the connection between the PH assumption and this family of models. % \ref{appendix:1}

Note that, for a fixed value $r$, the survival function $ S_r(t)$ also defines a Cure-Rate model. Indeed, $\lim_{t \rightarrow +\infty} S_r(t) = p^r$, such that $S_r(t)$ can be written as
\begin{eqnarray}
S_r(t) = p^r + \left( 1-p^r \right) \widetilde{S}_r(t),
\label{CR}
\end{eqnarray}
with conditional (proper) survival function and density function for susceptibles respectively equal to
\begin{eqnarray*}
&\widetilde{S}_r(t) = \dfrac{\left[ p + (1-p) \widetilde{S}(t) \right]^r - p^r}{1 - p^r},
&\widetilde{f}_r(t) 
= \dfrac{1 - p}{1 - p^r} \, r \left[ p + (1 - p) \widetilde{S}(t) \right]^{(r - 1)} \widetilde{f}(t).
\end{eqnarray*}
The proof of the obvious fact that a proper (conditional) density function does not exist for a Cure-Rate model is reported in Web Appendix A of Supplementary Materials.

We assume that the frailty risk $R$ follows a Gamma($\theta, \theta$) distribution with the shape and rate parametrization. This corresponds to the density function \begin{eqnarray}
    g_R(r;\theta) = \dfrac{\theta^\theta}{\Gamma(\theta)}r^{\theta-1}\text{e}^{-r\theta}, \ \theta > 0, \ r > 0.
    \label{prior-frailty}
\end{eqnarray}
Note that the Cure-Rate model (\ref{CR}), when coupled with the Gamma frailty (\ref{prior-frailty}), is such that the conditional cure-rate fraction $p^r$ corresponds to the marginal proportion of non-susceptibles $p_{\text{marg}} = P(\text{non-susceptible}) =  {\theta^\theta}/{(\theta-\log (p))^\theta}$. In details, 
\begin{eqnarray*}
&p_{\text{marg}} = \mathbb{E}_R(P(\text{non-susceptible}\mid R)) = \mathbb{E}_R(p^R) = \mathbb{E}_R(\text{e}^{\log(p^R)}) = \mathbb{E}_R(\text{e}^{R\log(p)}) =\text{MGF}_R(\log(p)) =\left(1 - \dfrac{\log(p)}{\theta} \right)^{-\theta} = \\
&= \left(\dfrac{\theta - \log(p)}{\theta} \right)^{-\theta} = \left(\dfrac{\theta}{\theta - \log(p)} \right)^{\theta},
\end{eqnarray*}
using the Moment Generating Function (MGF). Note that, as $\theta\to\infty$, $\lim_{\theta\to\infty}P_\text{marg} = p$, as expected since increasing $\theta$ corresponds to decreasing heterogeneity ($\text{var}(R)\to0$). 

Assuming data are drawn from the Multivariate Shared Frailty Cure-Rate model, the likelihood of $\theta$, $p$, and the parameters of the baseline survival distribution of susceptibles $\underline{\gamma}$, all collected in the parameter collection $\underline \pi = (\theta, p, \underline\gamma)$, is
\begin{eqnarray*}
\label{multivariatemodel}
    &L(\underline\pi;\underline X)=\prod_{i=1}^n\prod_{j=1}^{n_i}\left[\dfrac{(1-p)\widetilde f(x_{ij})}{p + (1-p)\widetilde S(x_{ij})}\right]^{\delta_{ij}} \dfrac{\Gamma(\theta + \sum_{j=1}^{n_i}\delta_{ij} )}{\Gamma(\theta)\theta^{\sum_{j=1}^{n_i}\delta_{ij}}}\left(1 - \dfrac{\log\left(\prod_{j = 1}^{n_i}\left(p + (1-p)\widetilde S(x_{ij})\right)\right)}{\theta}\right)^{-(\theta + \sum_{j=1}^{n_i}\delta_{ij})}
\end{eqnarray*} 
with $\boldsymbol{\delta}_i = (\delta_{i1}, \dots, \delta_{in_i})$.
This easily reduces to a simpler form in the univariate case. See (non-web) Appendix B for full details. % \ref{appendix:2}

\subsection{Family risk prediction}\label{subsec:2.2}
Our primary focus is risk prediction for women not included in the available dataset. Risk prediction in the parametric setting is achieved through the computation of a summary of the family-level frailty posterior distribution conditional on the whole history of the family: 
\begin{eqnarray*}
        g_R\left(r\mid \underline{X}_i;\theta\right)
        \sim\text{Gamma} \left(\theta + \sum_{j=1}^{n_i}\delta_{ij}, \theta -\sum_{j=1}^{n_i}\log\left(S_0\left(x_{ij}\right)\right)\right),
\end{eqnarray*} with the shape and rate parametrization. This posterior distribution is easily computed by relying on the conjugacy property of the Gamma prior for this model. The form for the univariate case can be straightforwardly derived. Details are shown in (non-web) Appendix C (see also for instance, \cite{tyrer2004breast, rosner2008risk, darabi2012breast}). % \ref{appendix:3}
From the posterior distribution, one can compute the posterior mean of a Gamma (see for instance, \cite{balan2019frailtyem})
\begin{eqnarray}
\label{formula:2}
    \widehat r_i = \dfrac{\widehat \theta + \sum_{j=1}^{n_i}\delta_{ij}}{\widehat \theta -\sum_{j=1}^{n_i}\log\left(\widehat p  + (1-\widehat p )\widehat S_0(x_{ij})\right)},
\end{eqnarray}
or the posterior median which, however, does not have a closed form and thus must be (easily) computed through numerical algorithms. We compare the posterior mean and median in terms of mean squared prediction error (MSPE) ${1}/{n}\sum_{i=1}^n( r_i$ $- \widehat r_i)^2$, correlation $\rho$ (and rank correlation $Rank \ \rho$), and coefficient of determination $R^2$ from a linear regression of the true ($r_i$) on the predicted risk ($\widehat r_i$), in order to elect the best to be used in the model comparative analysis.

We involved the Harrell's concordance index to measure the model coherence in risk prediction (see for instance, \cite{rahman2017review, harrell1982evaluating, scheike2015estimating, tokatli2011developing}). This choice is motivated by the fact that, in the real dataset, a direct comparison between the predicted frailty and the true frailty is not feasible.  This index  estimates the probability that, among a randomly selected pair of subjects with different right-censored survival times and predicted frailty values, the one with a lower predicted risk will outlive the other (see for instance, \cite{harrell1982evaluating}), as a measure of concordance between survival times and risks. Notice that this is similar to what is done in the modified version of the Wilcoxon test on survival data with right censoring. The estimate is based on the proportion of concordant pairs over all pairs (including discordant) that can be compared (see for instance, \cite{schmid2016use})
\[
C = \dfrac{\sum_{i\ne i', j \ne j'}\mathbb{I}(\widehat{r}_{ij} < \widehat{r}_{i'j'})\mathbb{I}(x_{ij} > x_{i'j'})\delta_{i'j'}}{\sum_{i\ne i', j \ne j'}\mathbb{I}(x_{ij} > x_{i'j'})\delta_{i'j'}},
\] where $\widehat{r}_{ij}$ and $\widehat{r}_{i'j'}$ denote the predicted frailty risk for two subjects $j$ and $j'$ who belong to two different families $i$ and $i'$, respectively. 
% Additional results on the Harrell's concordance can be found in Appendix D.

To validate model performance for binary risk classification, we first achieve risk prediction by computing the posterior probability that each family belongs to the highest-risk group - defined as those exceeding the upper $\alpha$ percentile of the prior risk distribution 
\begin{eqnarray*}
    \widehat{r}_i = G_R\left(G^{-1}_R\left(\alpha; \widehat{\theta}\right)\mid \underline X_i; \widehat{\theta}\right),
\end{eqnarray*} and then compute the Area Under the Receiver Operating Characteristic Curve (AUC), the Positive Predictive Value (PPV), and the Negative Predictive Value (NPV).

\section{Comparative analysis based on a simulation study}\label{section:3}
As an alternative to the Multivariate Shared Frailty Cure-Rate model, one can rely on the Cox PH model. Although the true underlying structure of this model may still be based on the LCR structure (in Formula \ref{formula:1}), one may think of a model that has a proper survival distribution, both conditionally on $R$ and marginally. Indeed, one may assume that there exists some finite time $T_\text{max}$ (larger than all observed survival and censoring times) such that all conditional survival functions $S(t\mid r) = \left[S_0(t) \right]^r$ are proper. Therefore, the semiparametric estimation of the model is possible
(see for instance, \cite{balan2019frailtyem, balan2020tutorial}), while maintaining the multiplicative frailty structure.
Thus, the Multivariate Shared Frailty Cox model is formulated in its conventional form based on the existing literature (see for instance, \cite{cox1972regression}), with the only difference that the proportion of non-susceptibles cannot be estimated. In particular, we aimed to evaluate the adequacy of the parametric specification by comparing the Multivariate Shared Frailty Cure-Rate model to the Cox model.

A Univariate Cure-Rate model that incorporates only the observed subject-specific family history ($FH$) indicator was also included in the analysis (see, full details in Appendix D). % \ref{appendix:4}
The indicator $FH$ assumes value zero until the first case of breast cancer is observed in the family, after which it takes value one. The observed $FH$ is used as the predicted risk. However, a binary risk model is unlikely to accurately capture the variability of the continuous risk. A comparison between $FH$ and the true frailty risk $R$ with data generated from the Multivariate Shared Frailty Cure-Rate model can be found in Web Appendix B. 

For sick of completeness, the comparative analysis also involved what we have called the Univariate Frailty Cure-Rate model (see, full details in Appendix E), % \ref{appendix:5}
and two Univariate Cox models which include the covariate $FH$ measured at two different time points: one at the end of follow-up ($FH$), and one over the follow-up period ($FH(t)$)—
tracking the change-point from zero to one at the time it occurs—.

\subsection{Data generation from the Multivariate Shared Frailty Cure-Rate model}

Survival times for susceptibles and non-susceptibles were generated separately (see for instance, \cite{mota2022new}). We referred to the model in Formula \ref{CR}. If the probability of being susceptible (randomly generated from a Uniform distribution) was greater than $p^r$, the survival time $t$ was \[
    t = \widetilde{S}^{-1} \left( \frac{u^{1/r} - p}{1-p} \right),
\] otherwise the subject was non-susceptible. Under the conditional independence assumption within each family, all family members had the same frailty risk $r$. The available functions in software \texttt{R} packages allowed us to easily compute quantiles from the distribution we set.

Birthdays were generated from a Uniform distribution, according to the generation. Mothers' birthdays were generated between 1905 and 1945, while daughters' birthdays were generated by adding to the mother's birthday a random number from a Uniform distribution between 25 and 35, which corresponds to the age at which the mother gave birth. If one takes into account that family members from different generations have different birthdays (hence in calendar time), follow-up is, clearly, longer for members born earlier. Administrative right censoring was generated as the minimum between death (which similarly follows the generation of birthdays), lost-at-follow-up time, and 2020, the year when follow-up ends. If a proper (non-cure rate) survival function were involved, then the data generation would be simpler. For example, if $S_r(t)=[S_0(t)]^r$ with $S_0(t)$ the traditional survival function of a Weibull($k,\lambda$) random variable, it is immediate to check that $T\mid R = r\sim$Weibull$\left( k, \lambda r^{1/k}\right)$, so that a simple reparametrization involving the frailty term $R$ would be enough to generate from the conditional distribution.

We conducted the analysis on $n=100,000$ families for $100$ repetitions to obtain parameter average point estimates and their standard errors. When moving from the multivariate to the univariate setting, we ensured a sufficiently large sample size of $100,000$ main subjects. Also, results suggest that a sample size of $100,000$ families is sufficient to obtain accurate results with reasonable computational speed. We randomly generated the number of female family members by adding one (to ensure to have families with at least one member) to the minimum between a fixed number ($n_F-1$) and a number generated from a Poisson with parameter $\lambda_F$. Hence, $n_F$ is the maximum number of female members, and $\lambda_F$ the average number (prior to truncation by $n_F$) of first-degree female relatives. 

\subsection{Simulation results}
The Multivariate Shared Frailty Cure-Rate model was initially fitted with several susceptible survival distributions in order to explore estimation stability. Parameters were set at $(n_F, \ \lambda_F) = (5, \ 0.8)$, and $p =0.85$, while $\theta$ was varying among $(0.2,\ 0.5,\ 0.8)$.  This analysis consisted of data generated under a single combination of parameter values, as our limited simulation experiment was designed to capture the qualitative behaviour of different models under a setting similar to that of the Swedish Multi-Generational Breast Cancer registry data (see, full details in Section \ref{section:4}).
Results from a Weibull($\text{shape}=8$, $\text{scale}=6$), a Gamma($\text{shape}  = 8,\ \text{scale} =6$), a Lognormal($\mu  = 8,\ \sigma =6$), and a 3-parameter Gamma($\text{shape} =8,\ \text{scale} =6,\ \gamma =15$) suggest that parameter estimation is accurate independently of the chosen survival distribution (see, full details in Table F.1, Appendix F). The model is identifiable. % \ref{table:2}, \ref{appendix:6}

The parametric survival distribution was set as a Weibull($\text{shape}=8$, $\text{scale}=6$). Results show the estimation accuracy of the Multivariate Shared Frailty Cure-Rate model and the Multivariate Shared Frailty Cox model, when data are generated from the Multivariate Shared Frailty Weibull Cure-Rate model (see, full details in Table F.2, Appendix F). The same cannot be stated for the univariate models where, specifically for the $FH$ models, the frailty parameter $\theta$ was replaced by $\beta$ which captures the risk difference between families with a negative family history ($FH = 0$), and a positive family history ($FH = 1$) (see, full details in Table F.3, Appendix F). A proper comparison to the true parameter values cannot be made for these models.

After parameter estimation, we computed the frailty posterior distribution for each family and compared it to the true frailty values to obtain a preliminary measure of prediction accuracy (see, Tables \ref{table:10} and \ref{table:11}), also identifying the best summary measure (mean or median) as described in Section \ref{subsec:2.2}. For the Cox model, unlike the parametric approach, the baseline survival function was estimated non-parametrically using the \cite{breslow1972discussion} method , and the predicted mean risk at the family level (Formula \ref{formula:2}) was obtained via an Empirical Bayes approach (see for instance, \cite{balan2020tutorial}), with linear interpolation applied when cumulative hazard values were unavailable (see for instance, \cite{meijering2002chronology}). This procedure was also used in the EM algorithm with a different Gamma parametrization (see for instance, \cite{duchateau2008frailty}). 

Several combinations of parameters $(n_f,\ \lambda_F)\in\{(2,\ 0.8),\ (5,\ 0.8),\ (10,\ 5),\ (20,\ 10)\}$ captured differences from small to large family sizes, with larger families generally improving risk prediction accuracy. 

Simulation results highlight clear differences in predictive performance between multivariate and univariate models across these scenarios. In multivariate models, the Multivariate Shared Frailty Cure-Rate model shows moderate MSPE and correlation for smaller frailty counts (e.g., MSPE=3.87, $\rho$=0.45 for (2,0.8), Table \ref{table:10}) but improves substantially as family size increases, reaching MSPE=3.43 and $\rho$=0.78 for (20,10). Using the posterior median slightly increases MSPE while maintaining similar correlations. The Multivariate Shared Frailty Cox model generally outperforms the Cure-Rate model when using the posterior mean, though the posterior median performs worse. The univariate models, by relying only on $FH$ or $FH(t)$ indicators, show consistently higher MSPE and lower correlations, indicating that ignoring complete familial information reduces predictive accuracy. Overall, these results emphasize that incorporating multivariate dependence and a cure-rate structure substantially enhances risk prediction, particularly in scenarios with larger and more variable families, and that the posterior mean is generally preferable for prediction. % \ref{table:10}, \ref{table:11}, \ref{appendix:6}
The only drawback of posterior mean is that, for small sample sizes, it exhibits a well-known shrinkage problem (see for instance, \cite{balan2020tutorial}) due to the low number of events within families, highlighting the need for caution regarding the number of female family members in the data. 

\begin{table}[bt]
\caption{Summary of simulation results in 100 studies, prediction accuracy of posterior mean and median as predicted risk in multivariate models: empirical mean (standard error of the mean) of MSPE, Pearson’s $\rho$, rank correlation, and $R^2$.}
\label{table:10}
\begin{threeparttable}
\begin{tabular}{lllccccc}
\headrow
&&&&\thead{($\boldsymbol{n}_F, \boldsymbol{\lambda}_F$)} && \\
% \arrayrulecolor{black}\cline{3-6}
\headrow
\thead{Model} &\thead{Frailty} &\thead{Index} &\thead{(2, 0.8)} &\thead{(5, 0.8)} &\thead{(10, 5)} &\thead{(20, 10)} \\
\hiderowcolors
Multivariate & Mean & MSPE & 3.87 (0.049) & 4.80 (0.074) & 3.84 (0.071) & 3.43 (0.049) \\
Shared Frailty && $R^2$ & 0.20 (0.002) & 0.22 (0.002) & 0.48 (0.004) & 0.61 (0.002) \\
Cure-Rate model && $\rho$ & 0.45 (0.002) & 0.47 (0.002) & 0.69 (0.007) & 0.78 (0.001) \\
&& Rank $\rho$ & 0.18 (0.002) & 0.23 (0.002) & 0.44 (0.002) & 0.54 (0.001) \\
&Median & MSPE & 4.42 (0.052) & 5.40 (0.078) & 4.18 (0.074) & 3.64 (0.051) \\
&& $R^2$ & 0.21 (0.002) & 0.22 (0.002) & 0.48 (0.004) & 0.62 (0.002) \\
&& $\rho$ & 0.45 (0.002) & 0.47 (0.002) & 0.69 (0.003) & 0.79 (0.001) \\
&& Rank $\rho$ & 0.18 (0.002) & 0.27 (0.002) & 0.44 (0.002) & 0.54 (0.001) \\
\showrowcolors
\rowcolor{black!10}
Multivariate &Mean & MSPE & 3.80 (0.002) & 3.50 (0.002) & 2.85 (0.002) & 3.63 (0.003) \\
\rowcolor{black!10}
Shared Frailty && $R^2$ & 0.23 (<0.001) & 0.26 (<0.001) & 0.52 (<0.001) & 0.65 (<0.001) \\
Cox model && $\rho$ & 0.49 (<0.001) & 0.55 (<0.001) & 0.74 (<0.001) & 0.82 (<0.001) \\
\rowcolor{black!10}
&& Rank $\rho$ & 0.24 (<0.001) & 0.28 (<0.001) & 0.48 (<0.001) & 0.58 (<0.001) \\
&Median & MSPE & 5.79 (<0.001) & 5.64 (<0.001) & 5.01 (<0.001) & 3.89 (<0.001) \\
\rowcolor{black!10}
&& $R^2$ & 0.16 (<0.001) & 0.16 (<0.001) & 0.26 (<0.001) & 0.29 (<0.001) \\
&& $\rho$ & 0.42 (<0.001) & 0.43 (<0.001) & 0.53 (<0.001) & 0.55 (<0.001) \\
\rowcolor{black!10}
&& Rank $\rho$ & 0.11 (<0.001) & 0.15 (<0.001) & 0.36 (<0.001) & 0.49 (<0.001) \\
\hline
\end{tabular}
\end{threeparttable}
\end{table}

\begin{table}
\caption{Summary of simulation results in 100 studies, prediction accuracy of posterior mean, posterior median and $FH$ indicator as predicted risk in univariate models: empirical mean (standard error of the mean) of MSPE, Pearson’s $\rho$, rank correlation, and $R^2$.}
\begin{threeparttable}
\begin{tabular}{lllccccc}
\headrow
&&&&\thead{($\boldsymbol{n}_F, \boldsymbol{\lambda}_F$)} && \\
% \arrayrulecolor{black}\cline{3-6}
\headrow
\thead{Model} &\thead{Frailty} &\thead{Index} &\thead{(2, 0.8)} &\thead{(5, 0.8)} &\thead{(10, 5)} &\thead{(20, 10)} \\
\rowcolor{white}
Univariate &Mean & MSPE & 4.24 (0.053) & 5.26 (0.081) & 4.74 (0.084) & 4.51 (0.059) \\
Frailty && $R^2$ & 0.14 (0.003) & 0.16 (0.003) & 0.18 (0.002) & 0.13 (0.002) \\
\rowcolor{white}
Cox model && $\rho$ & 0.36 (0.004) & 0.39 (0.003) & 0.42 (0.003) & 0.35 (0.003) \\
&& Rank $\rho$ & 0.20 (0.004) & 0.21 (0.002) & 0.19 (0.002) & 0.18 (0.002) \\
\rowcolor{white}
& Median & MSPE & 4.54 (0.055) & 5.59 (0.084) & 5.04 (0.088) & 4.80 (0.062) \\
&& $R^2$ & 0.14 (0.003) & 0.16 (0.003) & 0.19 (0.003) & 0.13 (0.002) \\
\rowcolor{white}
&& $\rho$ & 0.36 (0.004) & 0.39 (0.003) & 0.43 (0.003) & 0.36 (0.003) \\
&& Rank $\rho$ & 0.20 (0.004) & 0.21 (0.002) & 0.19 (0.002) & 0.18 (0.002) \\
Univariate &$FH$ &MSPE &5.39 (0.057) &6.57 (0.09) &5.62 (0.097) &5.03 (0.062) \\
\rowcolor{black!10}
$FH$ &&$R^2$ &0.04 (0.001) &0.04 (0.001) &0.15 (0.003) &0.19 (0.002) \\ 
Cure-Rate model &&$\rho$ &0.19 (0.004) &0.19 (0.003) &0.38 (0.004) &0.43 (0.002) \\ 
\rowcolor{black!10}
&&Rank $\rho$ &0.15 (0.002) &0.16 (0.002) &0.35 (0.001) &0.43 (0.001) \\ 
\rowcolor{white}
Univariate &$FH$ &MSPE &5.91 (0.002) &5.87 (0.004) &5.67 (0.003) &5.23 (0.003) \\ 
$FH$ &&$R^2$ &0.03 ($<0.001$) &0.04 ($<0.001$) &0.13 ($<0.001$) &0.16 ($<$0.001) \\
\rowcolor{white}
Cox &&$\rho$ &0.17 ($<0.001$) &0.19 ($<0.001$) &0.36 ($<0.001$) &0.40 ($<0.001$) \\ 
&&Rank $\rho$ &0.14 ($<0.001$) &0.16 ($<0.001$) &0.35 ($<0.001$) &0.43 ($<0.001$) \\ 
Univariate &$FH(t)$ &MSPE &2.32 (0.007) &2.30 (0.006) &2.08 (0.006) &2.22 (0.006) \\ 
\rowcolor{black!10}
$FH(t)$ &&$R^2$ &0.03 ($<0.001$) &0.04 ($<0.001$) &0.05 ($<0.001$) &0.04 ($<0.001$) \\ 
Cox &&$\rho$ &0.18 (0.001) &0.19 (0.001) &0.23 ($<0.001$) &0.19 ($<0.001$) \\ 
\rowcolor{black!10}
&&Rank $\rho$ &0.18 (0.001) &0.20 (0.001) &0.28 ($<0.001$) &0.25 ($<0.001$) \\ 
\hline
\end{tabular}
\end{threeparttable}
\label{table:11}
\end{table}

\begin{figure}
    \centering
    \includegraphics[width=\linewidth]{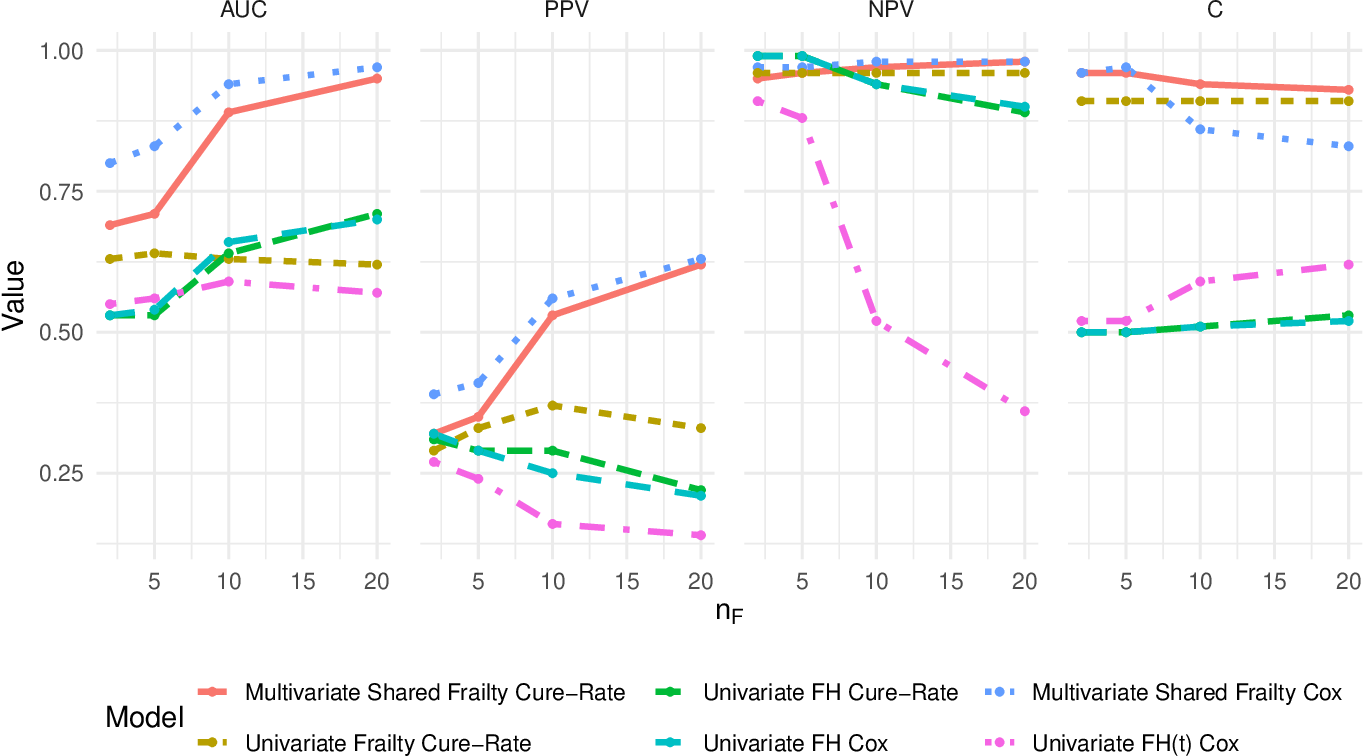}
    \caption{Model comparison in terms of AUC, Harrell's concordance index $C$, NPV, and PPV for  predicted versus true family-specific frailty values across different family sizes $(n_F, \lambda_F)$.}
    \label{figure:1}
\end{figure}

The posterior frailty mean was then used to evaluate risk prediction accuracy through AUC, PPV, and NPV for binary frailty risk, and Harrell's concordance index ($C$) for continuous risk. As illustrated in Figure \ref{figure:1} (see, first three panels), the Multivariate Shared Frailty Cure-Rate model (solid line) closely matches the Cox model (dotted line) in terms of binary risk (high-risk vs. low-risk families) prediction (see also, Table F.4, Appendix F where the Cure-Rate model achieves an average AUC = 0.95, PPV = 0.62, and NPV = 0.98 in larger families compared to the Cox model reaching in average AUC = 0.97, PPV = 0.63, and NPV = 0.98). The univariate models consistently underperform relative to multivariate models, with lower AUC and PPV (e.g., mean around 0.63 and 0.33 for the Univariate Frailty Cure-Rate model).

For continuous frailty risk instead, the Cure-Rate model outperforms the Cox model, with differences increasing as family size grows (see, last panel, Figure \ref{figure:1}). In more detail, the Multivariate Shared Frailty Cure-Rate model consistently achieves excellent concordance across all settings, maintaining high predictive accuracy (e.g., in average $C = 0.94$ for $(10,5)$ and $C = 0.93$ for $(20,10)$, Table F.4, Appendix F). By contrast, the Multivariate Shared Frailty Cox model exhibits a notable decline in concordance for larger families (e.g., in average $C = 0.86$ for $(10,5)$, and $0.83$ for $(20,10)$, Table F.4, Appendix F), highlighting the greater robustness of the Cure-Rate model in capturing continuous risk when more information is available. Moreover, the strong performance of the Univariate Frailty Cure-Rate model further supports the adequacy of the parametric assumptions, whereas the $FH$ models showed concordance values close to random prediction. 

All models, except the Univariate $FH(t)$ Cox, show - in average - a high NPV ($>0.89$) across settings, highlighting their common reliability in identifying low-risk families.

% figure_prediction.R in KI -> phd -> paper -> code

\section{Illustration to the Swedish Multi-Generational Breast \\ Cancer registry}\label{section:4}
This Section describes the application of all models to real data from the Swedish Multi-Generational Breast Cancer registry. Results are aligned with those observed in Section \ref{section:3}, supporting the conclusion that the Multivariate Shared Frailty Cure-Rate model provides the most accurate prediction of familial risk for invasive breast cancer detection.
\subsection{Data description}
The dataset we have analysed concerns a cohort of $n = 1,603,920$ Swedish families, consisting of a total of $4,267,803$ women. We excluded all men.

In each family, a woman randomly selected among those who were born between January 1, 1947 and December 31, 1976 was identified as the ``main subject.'' We chose these women because they were aged between 40 and 70 years at the end of the follow-up period (December 30, 2016). This is the age window associated with the highest probability of developing breast cancer. Identifying the main subject in each family allowed us to easily compare the multivariate models to the univariate models coherently by selecting always the same woman for each family.
The other family components were (if available) mother and sisters of the main subject. No distinction was made among them in this illustration (in particular, we did not account for birth cohort non-stationarity of survival). The 47.42\% of main subjects had no sisters, 35.55\% had one sister, and 11.85\% had two sisters (Table \ref{table:6}), making the majority of families consisting of two female members, followed by three and four members.
\begin{table}[ht]
\caption{Descriptive statistics of the Swedish Multi-Generational Breast Cancer registry.}
\label{table:6}
\begin{threeparttable}
\begin{tabular}{llll}
\headrow
\thead{Variable} & \thead{Min} & \thead{Median} & \thead{Max} \\
Birthday & 1853-11-15 & 1955-03-15 & 2018-12-15 \\
Diagnosis & 1958-01-15 & 1994-02-25 & 2016-12-30 \\
Death & 1947-05-08 & 1991-08-21 & 2018-12-31 \\
Emigration & 1948-01-01 & 1999-09-30 & 2018-12-31 \\
\hiderowcolors
\headrow
& \thead{Yes} & \thead{No} & \\
\rowcolor{white}
Breast cancer diagnosis &47,934 (2.99\%)&1,556,006 (97.01\%) & \\
\rowcolor{black!10}
Family history at end of follow-up &105,432 (6.57\%) &1,498,488 (93.43\%) & \\
\headrow
& \thead{Alive} & \thead{Censored} &\thead{Diagnosed} \\
\hiderowcolors
Status at end of follow-up & 1,408,072 (87.79\%) & 147,914 (9.22\%) &47,934 (2.99\%) \\
\rowcolor{black!20}
&\multicolumn{3}{l}{%
\textbf{0}\qquad\qquad
\textbf{1}\qquad\qquad
\textbf{2}\qquad\qquad
\textbf{3}\qquad\qquad
\textbf{4}\qquad\quad
\textbf{5}\qquad\quad
\textbf{6}\qquad\quad
\textbf{7}\qquad} \\
\rowcolor{white}
Number of sisters &\multicolumn{3}{l}{
47.42\%\quad
35.55\%\quad
11.85\%\quad
2.96\% \quad
0.98\% \quad
0.30\% \quad
0.89\% \quad
0.05\%\quad} \\
\hline
\end{tabular}
\end{threeparttable}
\end{table}

We excluded diagnoses of non-invasive ductal carcinoma in situ (DCIS), as our analysis focused only on the first detection of invasive breast cancer. However, to simplify the text, we refer to invasive breast cancer simply as ``breast cancer''. Time to breast cancer detection was right-censored by death, emigration, DCIS diagnosis, and by administrative censoring at the end of follow-up (December 30, 2016). Median age to breast cancer detection, computed by inverse censoring, was 53.29 years. Among all women, 47,934 (2.99\%) were diagnosed with invasive breast cancer, 147,914 (9.22\%) were censored before the end of the follow-up, and 1,408,072 (87.79\%) were alive at the end of the follow-up without having experienced breast cancer detection or any other censoring event (Table \ref{table:6}). In other words, 87.79\% is the proportion of observed non-susceptibles into the dataset, which suggests that a cure-rate survival structure may be appropriate. Indeed, this dataset provided a precious opportunity to explore the debated issue of whether incorporating a cure-rate structure would be appropriate in models for breast cancer detection. 

\begin{figure}[ht]
\centerline{\includegraphics[width=\textwidth]{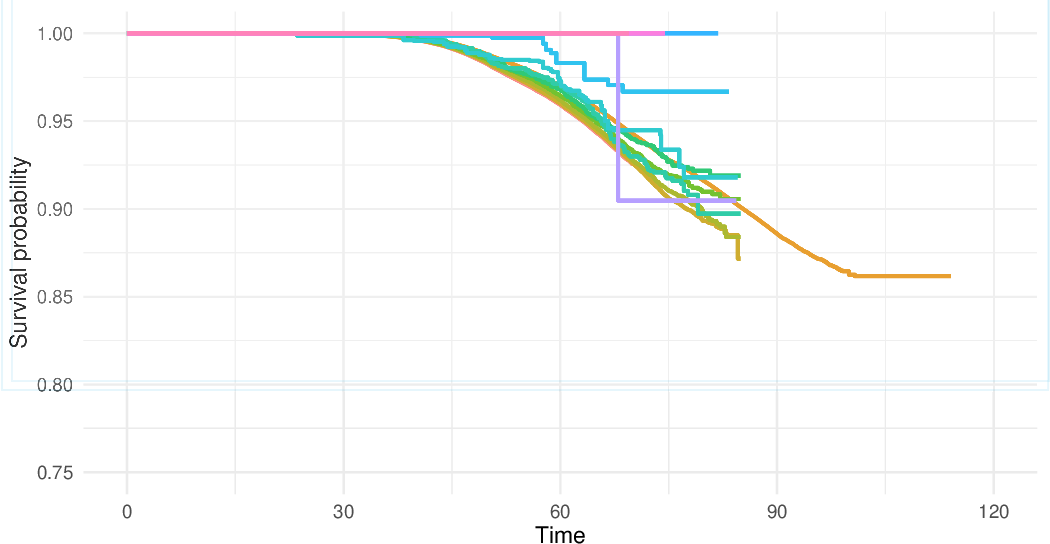}}
\caption{Kaplan–Meier survival curves for time to breast cancer diagnosis (years), shown separately for main subjects, other sisters and, most importantly, for mothers (longer tail). \label{figure:2}}
\end{figure}

Identifiability, which may be challenged by heavy censoring or limited follow-up (see for instance, \cite{hanin2014identifiability}), was ensured by the long follow-up in the Swedish Multi-Generational Breast Cancer registry. We verified the plausibility of the long survival tail, which was primarily driven by older mothers as shown in the Kaplan-Meier curve stratified by subjects in Figure \ref{figure:2} (see, longer tail), by comparing our ultracentenarian records to Swedish life tables (see, full details in Web Appendix C). This supports the reliability of the cure-rate estimate, consistent with identifiability results for mixture cure models under reasonable conditions (see for instance, \cite{li2001identifiability}). 

As an additional data quality control, we fitted a Univariate survival Cox model on the main subjects with a time-varying family history covariate $FH(t)$ on a nested case-control design, to estimate the overall relative risk associated with first-degree family history. We verified that our estimate, with value 1.8017, was consistent with the literature showing that the value is approximately 1.8 (see for instance, \cite{collaborative2001familial, pharoah1997family}).

\subsection{Implementation of the compared models}
We fitted the Multivariate Shared Frailty Cure-Rate model on the Swedish  Multi-Generational Breast Cancer registry. On average, the model estimates the frailty parameter $\theta$ to be approximately 5, which is higher than the estimate of 1.33 obtained from the Cox model (Table \ref{table:7}). Importantly, the Multivariate Shared Frailty Cure-Rate model yields a smaller residual heterogeneity across families (var($R$) = $1/\theta$), and allows for the direct estimation of the fraction of non-susceptibles which is, on average across the survival distributions, 85\% (Table \ref{table:7}). This probability refers to a family with $R=1$. Note that for $\widehat\theta \simeq 5$ and $\widehat p \simeq 0.85$ the marginal probability of being non-susceptible is close to the baseline value $p$. Indeed, (from Section \ref{section:2.1}) $p_{\text{marg}} = (\widehat\theta^{\widehat\theta})/(\widehat\theta-\log (\widehat p))^{\widehat\theta} = 5^5/(5 - \log(0.85))^5 = 0.8522$. 
\begin{table}[bt]
\caption{Summary of results from real data, mean (standard error of the mean) of estimated parameters of the Multivariate Shared Frailty Cure-Rate model with varying survival distributions among susceptibles (3p Gamma stands for 3-parameter Gamma); mean (standard error of the mean) of frailty parameter estimated through the Multivariate Shared Frailty Cox model.}
\begin{threeparttable}
\begin{tabular}{llcccccc}
\headrow
\thead{Model} &\thead{Distribution} &\thead{$\widehat{\boldsymbol{\theta}}$} &\thead{$\widehat{\boldsymbol{p}}$} &\thead{$\widehat{\text{shape}}/\widehat{\boldsymbol{\mu}}$} & \thead{$\widehat{\text{scale}}/\widehat{\boldsymbol{\sigma}}$} &\thead{$\widehat{\boldsymbol{\gamma}}$} \\
Cure-Rate &Weibull &4.36 (0.013) &0.87 (0.298) &6.18 (0.477) &72.98 (1.573) & \\ 
\rowcolor{white}
    &Gamma &5.48 (0.012) &0.85 (0.293) &18.69 (1.727) &3.86 (1.498) & \\ 
    &Lognormal &5.53 (0.012) &0.84 (0.288) &4.29 (1.492) &0.26 (0.590) & \\ 
\rowcolor{white}
    &3p Gamma &5.81 (0.011) &0.85 (0.310) &18.67 (1.766) &3.82 (1.545) &0.15 ($<0.001$) \\
\rowcolor{black!10}
Cox &&1.33 (0.0059) &&&&\\
\hline
\end{tabular}
%  \begin{tablenotes}
% \item CR, Cure-Rute. 
% \end{tablenotes}
\end{threeparttable}
\label{table:7}
\end{table}

The Multivariate Shared Frailty Cure-Rate model has a higher average risk prediction accuracy of 97\% compared to the Cox model (96\%) (see, full details in Table F.5, Appendix F). % \ref{table:9}, \ref{appendix:6}
Due to the limited follow-up and number of events among the main subjects, clearly convergence for the univariate models was hard (this was an additional advantage of fitting a multivariate model), and thus results were not reported because unstable over (susceptible) survival distributions. Moreover, they perform poorly in terms of prediction, with a concordance index closes to that expected by random chance. This is consistent with our observations in Section \ref{section:3}. 

Estimated parameters could be used then to perform 
out-of-sample posterior familial risk prediction: consider a woman whose mother has been detected with breast cancer at 90 years old, and a sister who is 45 years old. The output from the algorithm %(Algorithm \ref{algorithm:1}) 
indicates that the woman has a posterior frailty of 1.5762 (mean) or 2.7801 (median) and that she does not belong to the highest-risk families, with a probability of $0.1264 < 0.1512$, where 0.1512 is the estimated upper 5$th$ percentile of the prior risk distribution from real data.

\section{Discussion and conclusion}\label{section:5}
Breast cancer is one of the most common and challenging diseases in women, with family history being a key risk factor. Accurate risk prediction and early detection are essential to improve treatment success and survival.

In contrast to univariate models, we explored how the Multivariate Shared Frailty Cure-Rate model can provide deeper insights into disease detection and refine breast cancer risk assessments by modelling families as unit of analysis rather than individuals. Through simulation studies and real data from the Swedish Multi-Generational Breast Cancer registry, we proved that simplified family history summaries, such as the binary family history indicator used in univariate models, worsen the predictive performance when the data are generated from a realistic multivariate familial setting. 

Contrasting to the Cox model, the Multivariate Shared Frailty Cure-Rate model does not lose predictive performance - also it slightly outperforms the Cox model on average when the risk was continuous -, as measured by AUC and Harrell’s concordance index, and also estimates the non-susceptible fraction and survival curves for susceptibles, offering a broader understanding of breast cancer risk. We consider estimating the non-susceptible fraction as a realistic approach for the Swedish Multi-Generational Breast Cancer registry, where around the 85$\%$ of subjects have not experienced yet breast cancer detection within the observed follow-up. This result supports the cure-rate assumption in the context of breast cancer detection. Moreover, the cure-rate assumption has the advantage of enlarging the set of available models, within which the traditional (proper) PH survival model is nested through the constraint that $p = 0$. The identifiability of the non-susceptible fraction estimate was ensured by the long follow-up in the Swedish Multi-Generational Breast Cancer registry, whose the mothers contributed the most. These findings hold great promise for targeting high-risk families more effectively, enabling better screening and prevention strategies, and ultimately leading to earlier breast cancer detection and improved patient survival outcomes. 

One of the main challenges we faced was assessing the goodness-of-fit of Cure-Rate models in the presence of a large fraction of right-censored observations. Even basic graphical assessments of goodness-of-fit typically require substantial sample sizes. Fortunately, the size of the Swedish multi-generational breast cancer registry data made this feasible but computational heavy. 

Several developments could further enhance the effectiveness and applicability of the Multivariate Shared Frailty Cure-Rate model. An important direction for future work will be to extend the model to include covariates, either individual or familial. This can be achieved by modelling the cure fraction through a regression structure (e.g., a logistic link) and by incorporating risk factors multiplicatively in the hazard function for the non-cured individuals (for individual covariates), following a proportional hazards formulation. Such an extension would allow the model to quantify how individual and familial characteristics influence both the probability of being cured and the progression dynamics among those at risk. This can enhance precision in targeting and improve accuracy in identifying the frailty parameter, as well as classifying families into risk groups. 

We also aim to relax the assumption of identical survival distributions by introducing (say, polynomial) cohort effects, to distinguish between mother and sisters. This may capture the expected bias arising from the differential effect of screening on diagnoses over time and affect prognosis. 

A full assessment of the added value of the Multivariate Shared Frailty Cure-Rate model will clearly emerge when additional analyses can be conducted on other datasets. A key next step is comparing its prediction accuracy to the BOADICEA model, which is the gold standard at the moment, using data with family structure and genetic information. 

Methodological extensions could also involve exploring alternative frailty distributions, such as the Lognormal distribution (see for instance, \cite{duchateau2008frailty}), which may capture unobserved heterogeneity more flexibly than the commonly used Gamma frailty. In addition, Bayesian approaches (see for instance, \cite{karamoozian2021bayesian, lee2017differences}) offer a framework for incorporating prior information and quantifying uncertainty in parameter estimates, particularly in settings with sparse data or complex hierarchical structures, such as those arising from familial data.

\section*{Acknowledgements}
We would like to acknowledge Keith Humphreys for his precious advice and contribution. 

\section*{Conflict of interest}
The authors declare no potential conflict of interests.

\section*{Data availability}
The data underlying this article were provided by \textit{Socialstyrelsen} and Statistics Sweden by permission. Data is not allowed to be shared due to Swedish laws. However, it is possible to request the same data for research from Swedish registers.

\section*{Funding}
This work was supported by the Swedish Research Council (Grant No. 2022-00584) awarded to Kamila Czene. Marco Bonetti was supported in part by the MUSA – Multilayered Urban Sustainability Action – project, funded by the European Union – Next Generation EU, under the National Recovery and Resilience Plan (NRRP) Mission 4 Component 2 Investment Line 1.5.

\section*{Ethics approval}
The methods were carried out in accordance with the relevant guidelines and regulations. The study has received approval by the Bocconi Ethics Board (SA000483).

\section*{Code availability}
The R code developed and implemented for this work can be made available by contacting the first author.

\section*{Author contribution}
Maria Veronica Vinattieri and Marco Bonetti contributed equally to the technical developments. Maria Veronica Vinattieri implemented the methods and models. Kamila Czene provided access to the Swedish registry-based data and contributed expertise in the use of population-based data on family history of breast cancer. All three authors contributed equally to the writing of the manuscript.

\section*{Supplementary materials}
Web Appendices referenced in Sections
2, 3, and 4 are available with the
online version of the article at the publisher’s website.

% \printendnotes

% Submissions are not required to reflect the precise reference formatting of the journal (use of italics, bold etc.), however it is important that all key elements of each reference are included.
\clearpage
\bibliography{bibliography}

% \begin{biography}[example-image-1x1]{A.~One}
% Please check with the journal's author guidelines whether author biographies are required. They are usually only included for review-type articles, and typically require photos and brief biographies (up to 75 words) for each author.
% \bigskip
% \bigskip
% \end{biography}

% \graphicalabstract{example-image-1x1}{Please check the journal's author guildines for whether a graphical abstract, key points, new findings, or other items are required for display in the Table of Contents.}

\section*{Figure legend}
\begin{enumerate}
    \item[Figure 1] Model comparison in terms of AUC, Harrell's concordance index $C$, NPV, and PPV for  predicted versus true family-specific frailty values across different family sizes $(n_F, \lambda_F)$.
    \item[Figure 2] Kaplan–Meier survival curves for time to breast cancer diagnosis (years), shown separately for main subjects, other sisters and, most importantly, for mothers (longer tail).
\end{enumerate}

\end{document}

\author{Vinattieri \textsc{et al.}}
\section*{Appendix}

\begin{appendices}
\section{Lehmann family of Cure-Rate models and Proportional \\ Hazards}\label{appendix:1}
We now explore the meaning of the PH assumption when $P(T=+\infty)=p>0$. Indeed, in the Cure-Rate model we have $S_0(t) = p + (1-p) \widetilde{S}(t)$. A proportion $p$ of the population will never experience the event, no matter how long they will live, while the proportion $(1-p)$ are susceptibles to breast cancer detection according to the survival function $\widetilde S(t)$. 

\noindent
Let us compute the hazard function of a Cure-Rate model: \begin{eqnarray*}
    \lambda_0(t) &\overset{\Delta}{=}\lim_{\Delta_t\downarrow 0}\dfrac{1}{\Delta_t}P(T\in[t,t+\Delta_t]\mid T\ge t)=\lim_{\Delta_t\downarrow 0}\dfrac{1}{\Delta_t}\dfrac{P(T\in[t,t+\Delta_t])}{P(T\ge t)} \\
    &=\lim_{\Delta_t\downarrow 0}\dfrac{1}{\Delta_t}\dfrac{1}{S_0(t)}\left[F_0(t+\Delta_t)-F_0(t)\right]=\dfrac{1}{S_0(t)}\lim_{\Delta_t\downarrow 0}\dfrac{1}{\Delta_t}\left[S_0(t)-S_0(t+\Delta_t)\right] \\
    &=\dfrac{1}{S_0(t)}\lim_{\Delta_t\downarrow 0}\dfrac{1}{\Delta_t}\left[p+(1-p)\widetilde S(t)-(p+(1-p)\widetilde S(t+\Delta_t))\right] \\
    &=\dfrac{(1-p)}{S_0(t)}\lim_{\Delta_t\downarrow 0}\dfrac{1}{\Delta_t}\left[F_0(t+\Delta_t)-F_0(t)\right]=\dfrac{(1-p)}{S_0(t)}\widetilde f(t)=\dfrac{(1-p)\widetilde f(t)}{p+(1-p)\widetilde S(t)}.
\end{eqnarray*}

\noindent
Now, the PH assumption would require that for two groups A and B, one had (with obvious notation) \begin{eqnarray}
    \dfrac{(1-p_B)\widetilde f_{B}(t)}{p_B+(1-p_B)\widetilde S_{B}(t)} = \lambda_{B}(t) = \alpha\lambda_{A}(t)=\alpha\dfrac{(1-p_A)\widetilde f_{A}(t)}{p_A+(1-p_A)\widetilde S_{A}(t)}. 
    \label{formula:A}
\end{eqnarray} 
Note that this is different from the PH assumption on susceptibles. Indeed, consider the following cases: \begin{itemize}
    \item[(I)] if $p_A=p_B=0$ we recover the usual PH assumption on the susceptibles: $\widetilde\lambda_{B}(t)=\alpha\widetilde\lambda_{A}(t)$;
    \item[(II)] if $p_A=p_B=p>0$ and $\widetilde S_{A}(t)\ne \widetilde S_{B}(t)$ we have \begin{eqnarray*}
    \dfrac{\cancel{(1-p)}\widetilde f_{B}(t)}{p+(1-p)\widetilde S_{B}(t)} = \alpha\dfrac{\cancel{(1-p)}\widetilde f_{A}(t)}{p+(1-p)\widetilde S_{A}(t)};
    \end{eqnarray*}
    \item[(III)]  if $p_A, \ p_B>0$, $p_A\ne p_B$, and $\widetilde S_{A}(t)=\widetilde S_{B}(t)= \widetilde S(t)$ (so that $\widetilde f_{A}(t)= \widetilde f_{B}(t) = \widetilde f(t)$) we have \begin{eqnarray*}
        &\dfrac{(1-p_B)\widetilde f(t)}{p_B+(1-p_B)\widetilde S(t)} =\alpha\dfrac{(1-p_A)\widetilde f(t)}{p_A+(1-p_A)\widetilde S(t)} \\
        &\iff(1-p_B)(p_A+(1-p_A)\widetilde S(t))=\alpha(1-p_A)(p_B+(1-p_B)\widetilde S(t)) \\
        &\iff(1-p_B)(1-p_A)\widetilde S(t)+p_A(1-p_B) = \alpha(1-p_A)(1-p_B)\widetilde S(t)+\alpha p_B(1-p_A) \\
        &\iff \widetilde S(t) =\dfrac{\alpha (1-p_A)p_B-p_A(1-p_B)}{(1-\alpha)(1-p_A)(1-p_B)},
    \end{eqnarray*} i.e. it is constant, and the only case in which the survival function is constant is the degenerate case $\widetilde S(t)=1 \ \forall t$, (i.e. $P(T=+\infty)=1$).
    \item[(IV)] if $p_A=p_B=p>0$ and $\widetilde S_{A}(t)=\widetilde S_{B}(t)$ then, easily, $\alpha\equiv 1$;
    \item[(V)] if $p_A\ne p_B$ and $\widetilde S_{A}(t)\ne \widetilde S_{B}(t)$ then we maintain the general form in Formula \ref{formula:A} above. 
\end{itemize}

\section{Full likelihood of the Multivariate Gamma Shared Frailty Cure-Rate model}\label{appendix:2}
\noindent
The closed form of the multivariate likelihood $L(\theta, p, \underline\gamma;\underline X)$ for $i = 1,\dots, n$ families of varying size $n_i$ is
\begin{eqnarray*}
    &L(\theta, p, \underline\gamma;\underline X) = \prod_{i=1}^n\prod_{j=1}^{n_i}\int_{\mathbb {R}^+}f_r(x_{ij})^{\delta_{ij}}S_r(x_{ij})^{1 - \delta_{ij}}g_R(r;\theta)\text{d}r \\
    &=\prod_{i=1}^n\int_{\mathbb{R}^+}\prod_{j=1}^{n_i}\left[\dfrac{(1-p)\widetilde f(x_{ij})}{p + (1-p)\widetilde S(x_{ij})}r\cancel{(p + (1-p)\widetilde S(x_{ij}))^r}\right]^{\delta_{ij}}[p + (1-p)\widetilde S(x_{ij})]^{r(1-\cancel{\delta_{ij}})}g_R(r;\theta)\text{d}r \\
    &=\prod_{i=1}^n\prod_{j=1}^{n_i}\left[\dfrac{(1-p)\widetilde f(x_{ij})}{p + (1-p)\widetilde S(x_{ij})}\right]^{\delta_{ij}}\int_{\mathbb R^+}\prod_{j=1}^{n_i} r^{\delta_{ij}}S_r(x_{ij})g_R(r;\theta)\text{d}r \\
    &=\prod_{i=1}^n\prod_{j=1}^{n_i}\left[\dfrac{(1-p)\widetilde f(x_{ij})}{p + (1-p)\widetilde S(x_{ij})}\right]^{\delta_{ij}}\int_{\mathbb R^+}r^{\sum_{j=1}^{n_i}\delta_{ij}}\prod_{j=1}^{n_i}S_r(x_{ij})g_R(r;\theta)\text{d}r
\end{eqnarray*} 
Thus, given the general distribution $R\sim$Gamma$(\text{shape} = \alpha, \ \text{rate} = \beta)$, the development of the internal component is given by 
\begin{eqnarray*}
    &\int_{\mathbb R^+}r^{\sum_{j=1}^{n_i}\delta_{ij}}\prod_{j=1}^{n_i}S_r(x_{ij})g_R(r;\alpha,\beta)\text{d}r = \int_{\mathbb R^+}\prod_{j=1}^{n_i}S_r(x_{ij})r^{\sum_{j=1}^{n_i}\delta_{ij}}\dfrac{\beta^\alpha}{\Gamma(\alpha)}r^{\alpha-1}\text{e}^{-\beta r}\text{d}r \\ 
    &= \int_{\mathbb R^+}\prod_{j=1}^{n_i}S_r(x_{ij})\dfrac{\beta^{(\alpha + \sum_{j=1}^{n_i}\delta_{ij})}}{\Gamma(\alpha + \sum_{j=1}^{n_i}\delta_{ij})}\dfrac{\Gamma(\alpha + \sum_{j=1}^{n_i}\delta_{ij})}{\Gamma(\alpha)\beta^{\sum_{j=1}^{n_i}\delta_{ij}}}r^{(\alpha + \sum_{j=1}^{n_i}\delta_{ij}-1)}\text{e}^{-\beta r}\text{d}r \\ 
    &= \int_{\mathbb R^+}\prod_{j=1}^{n_i}S_r(x_{ij})\dfrac{\beta^{(\alpha + \sum_{j=1}^{n_i}\delta_{ij})}}{\Gamma(\alpha + \sum_{j=1}^{n_i}\delta_{ij})}\dfrac{\Gamma(\alpha + \sum_{j=1}^{n_i}\delta_{ij})}{\Gamma(\alpha)\beta^{\sum_{j=1}^{n_i}\delta_{ij}}}r^{(\alpha + \sum_{j=1}^{n_i}\delta_{ij}-1)}\text{e}^{-\beta r}\text{d}r \\ 
    &=\prod_{j=1}^{n_i}\dfrac{\Gamma(\alpha + \sum_{j=1}^{n_i}\delta_{ij})}{\Gamma(\alpha)\beta^{\sum_{j=1}^{n_i}\delta_{ij}}}\int_{\mathbb R^+}H(x_{ij};p,\underline\gamma)^rg_{R^*}(r;\alpha,\sum_{j=1}^{n_i}\delta_{ij},\beta)\text{d}r \\ 
    &=\prod_{j=1}^{n_i}\dfrac{\Gamma(\alpha + \sum_{j=1}^{n_i}\delta_{ij})}{\Gamma(\alpha)\beta^{\sum_{j=1}^{n_i}\delta_{ij}}} \int_{\mathbb R^+}
    \text{e}^{r\text{log}(H(x_{ij};p,\underline\gamma))}g_{R^*}(r;\alpha,\sum_{j=1}^{n_i}\delta_{ij},\beta)\text{d}r \\ 
    &=\prod_{j=1}^{n_i}\dfrac{\Gamma(\alpha + \sum_{j=1}^{n_i}\delta_{ij})}{\Gamma(\alpha)\beta^{\sum_{j=1}^{n_i}\delta_{ij}}} \mathbb{E}_{R^*}[\text{e}^{r\text{log}(H(x_{ij};p,\underline\gamma))}] \\
    &=\prod_{j=1}^{n_i}\dfrac{\Gamma(\alpha + \sum_{j=1}^{n_i}\delta_{ij})}{\Gamma(\alpha)\beta^{\sum_{j=1}^{n_i}\delta_{ij}}}\text{MGF}(R^*;\text{log}(H(x_{ij};p,\underline\gamma))) \\
    &=\prod_{j=1}^{n_i}\dfrac{\Gamma(\alpha + \sum_{j=1}^{n_i}\delta_{ij})}{\Gamma(\alpha)\beta^{\sum_{j=1}^{n_i}\delta_{ij}}}\left(1 - \dfrac{\text{log}(H(x_{ij};p,\underline\gamma))}{\beta}\right)^{-(\alpha + \sum_{j=1}^{n_i}\delta_{ij})}
\end{eqnarray*}
where we define the quantity $H(x_{ij};p,\underline\gamma) = \prod_{j = 1}^{n_i}S(x_{ij}) = \prod_{j = 1}^{n_i}\left(p + (1-p)\widetilde S(x_{ij})\right)$, with $\boldsymbol{\delta}_i = (\delta_{i1}, \dots, \delta_{in_i})$. The updated frailty distribution to obtain new frailty risk is $R^*\sim$Gamma$(\text{shape} = \alpha+\sum_{j=1}^{n_i}\delta_{ij},\text{rate}=\beta)$. The computation of the likelihood is possible thanks to the use of the MGF of $R^*$ in the point $\log(H(x_{ij};p,\underline\gamma))$. Recall that the definition of the MGF implies that \[
MGF(R;y) = \mathbb{E}_R[\text{e}^{ry}].
\] Specifically, for a Gamma distributed random variable $R\sim\text{Gamma}(\text{shape} = \alpha, \text{rate} = \beta)$ the MGF is given by \[
MGF(R;y) = \left(1-\dfrac{y}{\beta}\right)^{-\alpha}.
\]
The multivariate likelihood with $\alpha=\beta=\theta$ is given by \begin{eqnarray*}
\label{multivariatemodel}
    &L(\underline\pi;\underline X)=\prod_{i=1}^n\prod_{j=1}^{n_i}\left[\dfrac{(1-p)\widetilde f(x_{ij})}{p + (1-p)\widetilde S(x_{ij})}\right]^{\delta_{ij}}\dfrac{\Gamma(\theta + \sum_{j=1}^{n_i}\delta_{ij})}{\Gamma(\theta)\theta^{\sum_{j=1}^{n_i}\delta_{ij}}}\left(1 - \dfrac{\log\left(\prod_{j = 1}^{n_i}\left(p + (1-p)\widetilde S(x_{ij})\right)\right)}{\theta}\right)^{-(\theta + \sum_{j=1}^{n_i}\delta_{ij}) }
\end{eqnarray*} 
with $\underline \pi = (\theta, p, \underline\gamma)$. The likelihood reduces to the univariate form when there is one subject per family, following the formula \begin{eqnarray*}
    L(\underline\pi;\underline X)=\prod_{i=1}^n\left[\dfrac{(1-p)\widetilde f(x_{i1})}{p + (1-p)\widetilde S(x_{i1})}\right]^{\delta_{i1}}\dfrac{\Gamma(\theta + \sum_{j=1}^{n_i}\delta_{ij})}{\Gamma(\theta)\theta^{\sum_{j=1}^{n_i}\delta_{ij}}}\left(1 - \dfrac{\log\left(p + (1-p)\widetilde S(x_{i1})\right)}{\theta}\right)^{-(\theta + \sum_{j=1}^{n_i}\delta_{ij})}.
\end{eqnarray*}

\section{Posterior frailty risk distribution}\label{appendix:3}
Consider $H(x_{ij};p,\underline\gamma) = \prod_{j = 1}^{n_i}S(x_{ij}) = \prod_{j = 1}^{n_i}\left(p + (1-p)\widetilde S(x_{ij})\right)$, and define for the i$th$ family $f(\underline X_i\mid r) = \prod_{j=1}^{n_i}f(\underline x_{ij}\mid r) = \prod_{j=1}^{n_i}f_r(x_{ij})^{\delta_{ij}} S_r(x_{ij})^{1-\delta_{ij}}$.
\begin{eqnarray*}
    &g_R(r\mid \underline X_i;\theta) = \dfrac{f(\underline X_i\mid r_i)g_R(r)}{f(\underline X_i)} = \dfrac{\prod_{j=1}^{n_i}f(\underline x_{ij}\mid r)g_R(r)}{\int_{\mathbb{R}^+}\prod_{j=1}^{n_i}f(\underline x_{ij}\mid r)g_R(r)\text{d}r}=\\
    &=\dfrac{\prod_{j=1}^{n_i}\left(f_r(x_{ij})^{\delta_{ij}} S_r(x_{ij})^{1-\delta_{ij}}\right)g_R(r)}{\int_{\mathbb{R}^+}\prod_{j=1}^{n_i}\left(f_r(x_{ij})^{\delta_{ij}} S_r(x_{ij})^{1-\delta_{ij}}\right)g_R(r)\text{d}r} = \\
    &=\dfrac{\prod_{j=1}^{n_i}\left(p + (1-p)\widetilde S(x_{ij})\right)^r r^{\sum_{j=1}^{n_i}\delta_{ij}}g_R(r)}{\prod_{j=1}^{n_i}\dfrac{\Gamma(\theta + \sum_{j=1}^{n_i}\delta_{ij})}{\Gamma(\theta)\theta^{\sum_{j=1}^{n_i}\delta_{ij}}}}\left(1 - \dfrac{\text{log}(H(x_{ij};p,\underline\gamma))}{\theta}\right)^{(\theta + \sum_{j=1}^{n_i}\delta_{ij})}=\\
    &=\dfrac{\theta^{\sum_{j=1}^{n_i}\delta_{ij}}}{\Gamma(\theta + \sum_{j=1}^{n_i}\delta_{ij})}\left(1 - \dfrac{\text{log}\prod_{j=1}^{n_i}(p + (1-p)\widetilde S(x_{ij}))}{\theta}\right)^{(\theta + \sum_{j=1}^{n_i}\delta_{ij})}\prod_{j=1}^{n_i}\left(p + (1-p)\widetilde S(x_{ij})\right)^r r^{\sum_{j=1}^{n_i}\delta_{ij}}\theta^\theta r^{\theta-1}\text{e}^{-\theta r} \\
    &=\left(1 - \dfrac{\text{log}\prod_{j=1}^{n_i}(p + (1-p)\widetilde S(x_{ij}))}{\theta}\right)^{(\theta + \sum_{j=1}^{n_i}\delta_{ij})}\prod_{j=1}^{n_i}\left(p + (1-p)\widetilde S(x_j)\right)^r\dfrac{\theta^{(\theta + \sum_{j=1}^{n_i}\delta_{ij})}}{{\Gamma(\theta + \sum_{j=1}^{n_i}\delta_{ij})}}r^{(\theta + \sum_{j=1}^{n_i}\delta_{ij})-1}\text{e}^{-\theta r} \\
    &= \left(\theta-\log\prod_{j=1}^{n_i}(p + (1-p)\widetilde S(x_{ij}))\right)^{(\theta + \sum_{j=1}^{n_i}\delta_{ij})}\dfrac{1}{{\Gamma(\theta + \sum_{j=1}^{n_i}\delta_{ij})}}r^{(\theta + \sum_{j=1}^{n_i}\delta_{ij})-1}\text{e}^{-(\theta -\sum_{j=1}^{n_i}\log\left(p + (1-p)\widetilde S(x_{ij})\right))r} \\
    &\text{which is Gamma}\left(\text{shape} = \theta + \sum_{j=1}^{n_i}\delta_{ij}, \ \text{rate} = \theta -\sum_{j=1}^{n_i}\log\left(p + (1-p)\widetilde S(x_{ij})\right)\right).
\end{eqnarray*} 
\section{Observed data likelihood of the Univariate Family History Cure-Rate model}\label{appendix:4}
The Univariate $FH$ Cure-Rate model takes the form \begin{eqnarray*}
    &S_{FH}(t;fh) = [p+(1-p)\widetilde S(t)]^{fh(\beta-1)+1}, \ fh\in\{0, 1\}, \ \beta>0,
\end{eqnarray*}
or, using Formula 2 (see, Section 2 of the main text) % \ref{CR}, 
\begin{eqnarray}
    \label{cureratefh}
    S_{FH}(t;fh) = p^{fh(\beta-1)+1} + (1-p^{fh(\beta-1)+1})\widetilde S_{FH}(t;fh),
    \end{eqnarray}
with
\begin{eqnarray*}
    \widetilde S_{FH}(t;fh) = \dfrac{[p+(1-p)\widetilde S(t)]^{fh(\beta-1)+1} - p^{fh(\beta-1)+1}}{(1-p^{fh(\beta-1)+1})},
\end{eqnarray*}
and
\begin{eqnarray*}
    &\widetilde f_{FH}(t;fh) = \dfrac{(fh(\beta-1)+1)[p+(1-p)\widetilde S(t)]^{fh(\beta-1)} (1-p)\widetilde f(t)}{(1-p^{fh(\beta-1)+1})}.
\end{eqnarray*}
Notice that the frailty parameter $\theta$ is not involved in this model, due to the fact that it is replaced by the parameter $\beta$ which identifies the risk difference between the group of families with a negative family history ($fh = 0$), and those with a positive family history ($fh = 1$). When $fh = 0$ the survival function reduces to the baseline survival function $S_{FH}(t;fh=0) = [p+(1-p)\widetilde S(t)]$, otherwise $S_{FH}(t;fh=1) = [p+(1-p)\widetilde S(t)]^\beta$, which is also a Cure-Rate model (Formula \ref{cureratefh}). Thus, the Univariate $FH$ Cure-Rate model with the covariate family history indicator has likelihood \begin{eqnarray*}
    &L(\underline\pi_{FH};\underline X)=\dfrac{\prod_{i:fh_i = 0}\left[(1-p)\widetilde f_{FH}(t_i;0)\right]^{\delta_i} \left[p + (1-p)\widetilde S_{FH}(t_i;0)\right]^{1-\delta_i}}{\left(\prod_{i:fh_i = 1}\left[(1-p^\beta)\widetilde f_{FH}(t_i;1)\right]^{\delta_i} \left[p^\beta+(1-p^\beta)\widetilde S_{FH}(t_i;1)\right]^{1-\delta_i}\right)^{-1}}, 
\end{eqnarray*} with parameter collection $\underline\pi_{FH}=(\beta, p, \underline\gamma)^T$. 
\section{Observed data likelihood of the Univariate Frailty Cure-Rate model}\label{appendix:5}
Recall the (simplified) usual notation $X=\min(T,C)$ and $\Delta=\mathbb{I}(T \leq C)$ for the bivariate observed random variable arising from survival data $T$ independently right censored by the random variable $C$. It is easy to check that when $(T,C)$ has joint density function $f_{(T,C)}(t,c)=f_T(t) f_C(c)$, the distribution of $(X,\Delta)$ is proportional to $\left[f_T(x) \right]^\delta \left[S_T(x) \right]^{1-\delta}$. When the pairs $(T_i, C_i)$ are {\em i.i.d.} for $i=1, \ldots, n$, the product of such terms represents the observed data likelihood that can be maximized to learn about the distribution $F_T(t)$ ($F_C(c)$ is typically not of interest). In the following, $C$ is still assumed to be independent of $T$.

Now, consider the Cure-Rate model $S(t)=p + (1-p) \widetilde{S}(t)$, with $\widetilde{f}(t)$ the (proper) conditional density function of the time-to-event random variable for the susceptible subjects. Notice that $T$ has a positive probability $p$ of being equal to $+\infty$ (or to an extremely large number, as this model is sometimes also described). For ease of notation, below we write ``$\infty$'' for ``$+ \infty$.''

{\bf Proposition A}\label{proposition:1}
For the Cure-Rate model $S(t)=p + (1-p) \widetilde{S}(t)$, the contribution to the observed data likelihood by one observation $(X,\Delta)$ is proportional to the quantity $\left[ (1-p) \widetilde{f}(x) \right]^{\delta} \left[ p + (1-p) \widetilde{S}(x) \right]^{1-\delta}$.

\begin{proof}
Consider the probability $P(X \in [x, x+\Delta x), \Delta=0)$ for a non-negative, finite $x$. Define the set $A_T(x) = \{ (t,c) \in \mathbb{R}^+ \!\! \times \mathbb{R}^+ : c \in [x, x+\Delta x), t \geq c \}$. We have
\begin{eqnarray*}
  &P(X \in [x, x+\Delta x), \Delta=0) = P((T,C) \in A_T(x)) = \\
  & = P((T,C) \in A_T(x) \mid T < \infty) P(T < \infty)  + P((T,C) \in A_T(x) \mid T = \infty) P(T = \infty).
\end{eqnarray*}

It is easy to check that conditionally on $T < \infty$, $T$ and $C$ remain independent, with joint density function $f_{(T,C) \mid T<\infty}(t,c) = \widetilde{f}(t) f_C(c)$ on $\mathbb{R}^+ \!\! \times \mathbb{R}^+$. Therefore,
\begin{eqnarray*}
&P(X \in [x, x+\Delta x), \Delta=0) =  (1-p) \int_x^{x + \Delta x}   \int_u^{\infty} \widetilde{f} (t) f_C(c)  \text{d}t \,\text{d}c + p \int_x^{x+\Delta x} f_C(c)\text{d}c = \\
&= (1-p) \int_x^{x+\Delta x}f_C(c)  \widetilde{S}(c)\text{d}c + p \int_x^{x+\Delta x} f_C(c)\text{d}c\approx\\
&\approx  (1-p) (\Delta x) f_C(x) \widetilde{S}(x) + p \, (\Delta x) f_C(x) = \\
&= (\Delta x) \left[  f_C(x) \left( p + (1-p) \widetilde{S}(x) \right) \right].
\end{eqnarray*}

Now, define the set $A_C(x) = \{ (t,c) \in \mathbb{R}^+ \times \mathbb{R}^+ : t \in [x, x+\Delta x), c \geq t \}$. For $\Delta = 1$, slightly different steps yield
\begin{eqnarray*}
&P(X \in [x, x+\Delta x), \Delta=1) = P((T,C) \in A_C(x)) = \\
& =  P((T,C) \in A_C(x) \mid T < \infty) P(T < \infty)  + P((T,C) \in A_C(x) \mid T = \infty) P(T = \infty)  = \\
& = (1-p) \int_x^{x+\Delta x} \int_t^{\infty} f_C(c) \widetilde f(t)\text{d}c, \text{d}t + 0 = (1-p) \cdot \\
&\cdot \int_x^{x+\Delta x} \widetilde f(t) S_C(t) \text{d}t \approx  (\Delta x) (1-p) \widetilde f(x) S_C(x).
\end{eqnarray*}

Dividing by $\Delta x$, letting $\Delta x \rightarrow 0$, and writing the two terms in compact form produces the contribution
\begin{eqnarray*}
&\left[f_C(x) \left( p + (1-p) \widetilde S(x) \right) \right]^\delta \left[ (1-p) \widetilde f(x) S_C(x) \right]^{1-\delta} \\
&= \left( p + (1-p) \widetilde S(x) \right)^\delta \left[ (1-p) \widetilde f(x) \right]^{1-\delta}  \left[  f_C(x) \right]^\delta \left[ S_C(x) \right]^{1-\delta}\\
&\propto \left( p + (1-p) \widetilde S(x) \right)^\delta \left[ (1-p) \widetilde f(x) \right]^{1-\delta}.
\end{eqnarray*}
\end{proof}

Let us now turn to the LCR model structure.
\newline \ \newline
{\bf Proposition B} If $S_r(t) = S(t \mid R=r) = \left[ p + (1-p) \widetilde{S}(t) \right]^r$ $(r > 0)$, the contribution to the observed data likelihood provided by one observation $(X,\Delta)$ is proportional to the quantity
\[
\left[\dfrac{(1-p) \widetilde{f}(x)}{p + (1-p) \widetilde{S}(x)}  \right]^\delta  S_r(x) \, r^\delta.
\]

\begin{proof}
From earlier results, we can write $S_r(t) = \left[ p + (1-p) \widetilde{S}(t)  \right]^r = p^r + (1-p^r) \widetilde{S}_r(t)$, for
\[
\widetilde{S}_r(t) = \dfrac{\left[ p + (1-p) \widetilde{S}(t)  \right]^r - p^r}{1-p^r},
\]
and
\[
\widetilde{f}_r(t) = \dfrac{1-p}{1-p^r} r \left( p + (1-p) \widetilde{S}(t)  \right)^{r-1} \widetilde{f}(t).
\]
One can then use Proposition A for this new Cure-Rate model, replacing $p$ by $p^r$, $S(t)$ by $S_r(t)$, and $\widetilde{f}(t)$ by $\widetilde{f}_r(t)$. Simple algebra then yields the result.
\end{proof}
\section{Additional output from simulation studies and real data from the Swedish Multi-Generational Breast cancer registry}\label{appendix:6}

Table \ref{table:2} reports the summary of the parameter estimation accuracy of the Multivariate Shared Frailty Cure-Rate model under different survival distributions, based on 100 simulated datasets. For each distribution, we report the empirical mean and standard error of the estimated frailty parameter $\widehat{\boldsymbol{\theta}}$, the probability of being non-susceptible $\widehat{\boldsymbol{p}}$, and the parameters of the baseline survival distribution of susceptibles. 

Table \ref{table:1.1} and Table \ref{table:1.2} reports the summary of the simulation results across 100 studies for the Multivariate Shared Frailty Cure-Rate model, the Multivariate Shared Frailty Cox model, and the Univariate Frailty Cure-Rate model (Table \ref{table:1.1}), and the Univariate $FH$ Cure-Rate model, the Univariate $FH$ Cox model, and the Univariate $FH(t)$ model (Table \ref{table:1.2}). For each model, we report the empirical mean and standard error of the estimated frailty parameter $\widehat{\boldsymbol{\theta}}$, - or the family history parameter $\widehat{\boldsymbol{\beta}}$ - , the probability of being non-susceptible $\widehat{\boldsymbol{p}}$, and the parameters of the Weibull baseline survival distribution of susceptibles ($\widehat{\text{shape}}$ and $\widehat{\text{scale}}$).  

Table \ref{table:4} reports a summary of the model prediction accuracy in terms of Harrell's concordance index ($C$), Area Under the ROC curve (AUC), Positive Predictive Value (PPV), and Negative Predictive Value (NPV), under various combinations of $(n_f,\ \lambda_F)\in\{(2,\ 0.8),\ (5,\ 0.8),\ (10,\ 5),\ (20,\ 10)\}$. Results are shown for the posterior mean of the frailty-based predicted risk, as well as for the $FH$ indicator, over 100 simulated datasets. 

Table \ref{table:9} presents the estimated Harrell’s concordance index $C$, summarizing the predictive accuracy of various modelling approaches applied to real data from the Swedish Multi-Generational Breast Cancer registry. The comparison includes multivariate and univariate models with different survival distributions and frailty specifications.

\begin{table}[bt]
\caption{Summary of simulation results in 100 studies, empirical mean (standard error of the mean) of estimated parameters of the Multivariate Shared Frailty Cure-Rate model with varying susceptible survival distributions.}
\begin{threeparttable}
\begin{tabular}{lccccc}
\headrow
\thead{Distribution} &\thead{$\widehat{\boldsymbol{\theta}}$} &\thead{$\widehat{\boldsymbol{p}}$} &\thead{$\widehat{\text{shape}}/ \widehat{\boldsymbol{\mu}}$}& \thead{$\widehat{\text{scale}}/\widehat{\boldsymbol{\sigma}}$} &\thead{$\widehat{\boldsymbol{\gamma}}$} \\
     Weibull &0.20 (0.004) &0.85 (0.002) &7.99 (0.050) &5.99 (0.006) & \\
     \rowcolor{white}
     &0.50 (0.013) &0.85 (0.001) &8.00 (0.037) &6.00 (0.005) \\
     &0.80 (0.029) &0.85 (0.001) &8.00 (0.043) & 6.00 (0.006) \\
     Gamma &0.22 (0.074) &0.85 (1.123) &8.24 (1.013) &5.96 (0.096) & \\
     \rowcolor{black!10}
     &0.50 (0.022) &0.85 (0.002) &8.01 (0.097) &5.99 (0.085) & \\
     &0.79 (0.042) &0.85 (0.002) &7.99 (0.094) &6.01 (0.081) & \\
     Lognormal
     &0.19 (<0.001) &0.84 (0.001) &7.89 (0.019) &5.93 (0.006) & \\
     \rowcolor{white}
     &0.55 (0.002) &0.83 (0.001) &8.28 (0.024) &6.01 (0.007) & \\
    &0.95 (0.004) &0.84 (0.001) & 8.05 (0.023) & 6.01 (0.007) & \\
    3-parameter Gamma
     &0.19 (0.002) &0.84 (0.001) &9.34 (0.175) &6.85 (0.086) & 13.46 (0.399) \\ 
     \rowcolor{black!10}
     &0.61 (0.012) &0.84 (0.001) &7.99 (0.164) &7.64 (0.115) &17.32 (0.321) \\ 
     &0.88 (0.019) &0.86 (0.001) &8.06 (0.182) &6.48 (0.261) &17.56 (0.351) \\
    \hline
    \end{tabular}
\end{threeparttable}
\label{table:2}
\end{table}

\begin{table}[ht]
\caption{Summary of simulation results in 100 studies, empirical mean (standard error of the mean) of the estimated parameters of Multivariate Shared Frailty Cure-Rate model; Multivariate Shared Frailty Cox model; and Univariate Frailty Cure-Rate model.}
\begin{threeparttable}
\begin{tabular}{lcccc}
\headrow
\thead{Model} &\thead{$\widehat{\boldsymbol{\theta}}$} &\thead{$\widehat{\boldsymbol{p}}$} &\thead{$\widehat{\text{shape}}$} &\thead{$\widehat{\text{scale}}$} \\ 
     Multivariate Shared Frailty Cure-Rate model &0.20 (0.004) &0.85 (0.002) &7.99 (0.050) &5.99 (0.006) \\
     \hiderowcolors
     &0.50 (0.013) &0.85 (0.001) &8.00 (0.037) &6.00 (0.005) \\
     &0.80 (0.029) &0.85 (0.001) &8.00 (0.043) & 6.00 (0.006) \\ 
     \showrowcolors
     Multivariate Shared Frailty Cox model &0.19 (0.004) &-&-&- \\
     \rowcolor{black!10}
     &0.49 (0.012) &-&-&- \\ 
     &0.80 (0.029) &-&-&- \\ 
     \hiderowcolors
     Univariate Frailty Cure-Rate model &0.11 (0.001) &0.79 (0.001) &8.12 (0.037) &6.37 (0.037) \\
     &0.39 (0.017) &0.81 (0.001) &8.59 (0.024) &6.06 (0.003) \\
     &0.62 (0.057) &0.81 (0.001) &8.36 (0.025) &6.07 (0.003) \\
\hline
\end{tabular}
\end{threeparttable}
\label{table:1.1}
\end{table}

\begin{table}[bt]
\caption{Summary of simulation results in 100 studies, empirical mean (standard error of the mean) of estimated parameters of Univariate $FH$ Cure-Rate model; Univariate $FH$ Cox model; and Univariate $FH(t)$ Cox model.}
\begin{threeparttable}
\begin{tabular}{lcccc}
\headrow
\thead{Model} &\thead{$\widehat{\boldsymbol{\beta}}$} &\thead{$\widehat{\boldsymbol{p}}$} &\thead{$\widehat{\text{shape}}$} &\thead{$\widehat{\text{scale}}$} \\ 
     \hiderowcolors
     Univariate $FH$ Cure-Rate model
     &2.88 (0.059) &0.89 (0.001) &5.06 (0.065) &6.15 (0.084) \\
     &2.02 (0.050) &0.88 (0.002) &4.97 (0.057) &5.90 (0.012) \\
     &1.84 (0.038) &0.89 (0.001) &4.72 (0.063) &5.85 (0.015) \\ 
     \showrowcolors
     Univariate $FH$ Cox model
     &2.83 (0.081) &-&-&- \\ 
     \rowcolor{black!10}
     &1.45 (0.040) &-&-&- \\
     &1.10 (0.033) &-&-&- \\ 
     \hiderowcolors
     Univariate $FH(t)$ Cox model
     &5.85 (0.166) &-&-&- \\ 
     &2.96 (0.083) &-&-&- \\ 
     &2.23 (0.066) &-&-&- \\
\hline
\end{tabular}
\end{threeparttable}
\label{table:1.2}
\end{table}

\begin{table}
\caption{Summary of simulation results in 100 studies, empirical mean (standard error of the mean) of Harrell's concordance index $C$ if the risk is modelled as continuous; empirical mean (standard error of the mean) of AUC, PPV, and NPV if the risk is modelled as binary.}
\begin{threeparttable}
\begin{tabular}{llccccc}
\headrow
&&&\thead{($\boldsymbol{n}_F, \boldsymbol{\lambda}_F$)} && \\
% \arrayrulecolor{black}\cline{3-6}
\headrow
\thead{Model} &\thead{Index} &\thead{(2, 0.8)} &\thead{(5, 0.8)} &\thead{(10, 5)} &\thead{(20, 10)} \\
% \arrayrulecolor{black}\cline{3-6}
\hiderowcolors
    Multivariate Shared Frailty 
    Cure-Rate &$C$&0.96 ($<0.001$) &0.96 ($<0.001$) &0.94 (0.001) &0.93 ($<0.001$) \\
     model &AUC &0.69 (0.003) &0.71 (0.001)	&0.89 ($<0.001$) &0.95 (0.001) \\
    &PPV &0.32 (0.005) &0.35 (0.001) &0.53 (0.005) &0.62 (0.001) \\
    &NPV &0.95 (0.001) &0.96 ($<0.001$)	&0.97 ($<0.001$) &0.98 ($<0.001$) \\
\showrowcolors
    Multivariate Shared Frailty Cox model
    &$C$&0.96 ($<0.001$) &0.97 ($<0.001$) & 0.86 ($<0.001$) & 0.83 ($<0.001$) \\
    \rowcolor{black!10}
    &AUC &0.80 ($<0.001$) &0.83 ($<0.001$) &0.94 ($<0.001$) &0.97 ($<0.001$) \\
    &PPV &0.39 ($<0.001$) &0.41 ($<0.001$) &0.56 ($<0.001$) &0.63 ($<0.001$) \\
    \rowcolor{black!10}
    &NPV &0.97 ($<0.001$) &0.97 ($<0.001$) &0.98 ($<0.001$) &0.98 ($<0.001$) \\
\hiderowcolors
    Univariate Frailty Cure-Rate model
    &$C$&0.91 ($<0.001$) &0.91 ($<0.001$) &0.91 (0.001) &0.91 ($<0.001$) \\
    &AUC &0.63 (0.001) &0.64 (0.002) &0.63 (0.002) &0.62 ($<0.001$) \\
    &PPV &0.29 (0.005) &0.33 (0.002) &0.37 (0.006) &0.33 (0.002) \\ 
    &NPV &0.96 (0.001) &0.96 ($<0.001$)	&0.96 ($<0.001$) &0.96 ($<0.001$) \\
\showrowcolors
    Univariate $FH$ Cure-Rate model
     &$C$&0.50 ($<0.001$) & 0.50 ($<0.001$) &0.51 ($<0.001$) & 0.53 ($<0.001$) \\ 
     \rowcolor{black!10}
     &AUC &0.53 ($<0.001$) &0.53 ($<0.001$)	&0.64 ($<0.001$) &0.71 (0.001) \\
    &PPV &0.31 (0.010) &0.29 (0.009) &0.29 (0.002) &0.22 ($<0.001$) \\
    \rowcolor{black!10}
    &NPV &0.99 ($<0.001$) &0.99 ($<0.001$) &0.94 ($<0.001$)	&0.89 ($<0.001$) \\
\hiderowcolors
    Univariate $FH$ Cox model
    &$C$&0.50 ($<0.001$) & 0.50 ($<0.001$) &0.51 ($<0.001$) & 0.52 ($<0.001$) \\
    &AUC  &0.53 ($<0.001$) &0.54 ($<0.001$) &0.66 ($<0.001$) &0.70 ($<0.001$) \\
   &PPV &0.32 ($<0.001$) &0.29 ($<0.001$) &0.25 ($<0.001$) &0.21 ($<0.001$) \\
    &NPV &0.99 ($<0.001$) &0.99 ($<0.001$) &0.94 ($<0.001$) &0.90 ($<0.001$) \\ 
\showrowcolors
    Univariate $FH(t)$ Cox model
    &$C$& 0.52 ($<0.001$) & 0.52 ($<0.001$) & 0.59 ($<0.001$) & 0.62 ($<0.001$) \\
    \rowcolor{black!10}
    &AUC &0.55 ($<0.001$) &0.56 ($<0.001$) &0.59 ($<0.001$) &0.57 ($<0.001$) \\ 
    &PPV &0.27 (0.001) &0.24 (0.001) &0.16 ($<0.001$) &0.14 ($<0.001$) \\ 
    \rowcolor{black!10} 
    &NPV &0.91 ($<0.001$) &0.88 ($<0.001$) &0.52 (0.001) &0.36 ($<0.001$) \\ 
    \hline
    \end{tabular}
\end{threeparttable}
\label{table:4}
\end{table} 

\begin{table}[bt]
\caption{Summary of results from real data, mean (standard error of the mean) of the Harrell’s Concordance index $C$ of multivariate and univariate models, with varying susceptible survival distributions in Cure-Rate models.}
\begin{threeparttable}
\begin{tabular}{llc}%
    \headrow
    \thead{Model} &\thead{Distribution} &\thead{Harrell's Concordance index} \\
    Multivariate Shared Frailty Cure-Rate model &Weibull &0.97 ($< 0.001$) \\  
    \rowcolor{white}
    &Gamma &0.97 ($< 0.001$) \\ 
    &Lognormal &0.94 ($< 0.001$) \\ 
    \rowcolor{white}
    &3-parameter Gamma &0.97 ($< 0.001$) \\ 
    \rowcolor{black!10}
    Multivariate Shared Frailty Cox model &&0.96 ($< 0.001$)
    \\
    \rowcolor{white}
    Univariate Frailty Cure-Rate model &Weibull &0.39 ($< 0.001$) \\ 
    &Gamma &0.39 ($< 0.001$) \\ 
    \rowcolor{white}
    &Lognormal &0.51 ($< 0.001$) \\
    &3-parameter Gamma &0.45 ($< 0.001$) \\
    Univariate $FH$ Cure-Rate model &Weibull &0.50 ($< 0.001$) \\
    \rowcolor{black!10}
    &Gamma &0.51 ($< 0.001$) \\
    &Lognormal &0.47 ($< 0.001$) \\
    \rowcolor{black!10}
    &3-parameter Gamma &0.51 ($< 0.001$) \\
    \rowcolor{white}
    Univariate $FH$ Cox model &&0.51 ($< 0.001$) \\
    \rowcolor{black!10}
    Univariate $FH(t)$ Cox model &&0.50 ($< 0.001$) \\
    \hline
\end{tabular}
\end{threeparttable}
\label{table:9}
\end{table}
\end{appendices}